\documentclass[twocolumn, showpacs,aps, prb]{revtex4}
\usepackage{graphicx}
\newcommand{\bq}{\begin{equation}}
\newcommand{\eq}{\end{equation}}
\newcommand{\bn}{\begin{eqnarray}}
\newcommand{\en}{\end{eqnarray}}

\begin{document}
\title{Finite frequency current fluctuations and the self-consistent perturbation theory for electron transport through quantum dot}
\author {Guo-Hui Ding}
\affiliation{Key Laboratory of Artificial Structures and Quantum Control (Ministry of Education),
 Department of Physics, Shanghai Jiao Tong University, Shanghai, 200240, China}
\author {Bing Dong}
\affiliation{Key Laboratory of Artificial Structures and Quantum Control (Ministry of Education),
 Department of Physics, Shanghai Jiao Tong University, Shanghai, 200240, China}

\date{\today }

\begin{abstract}
 We have formulated the problem of electron transport through interacting quantum dot system in the framework of self-consistent perturbation theory, and show that the current conservation condition is guaranteed due to the gauge invariant properties of the Green's functions and the generalized Ward identity.  By using a generating functional for the statistics of the nonequilibrium system, we have obtained general formulae for calculating the current and the current fluctuations in the presence of arbitrary time-dependent potentials. As demonstration of application, we have studied the interaction effects on the finite frequency noise for electron resonant tunneling through an Anderson impurity, and obtained an analytical equation for the interaction effect on the finite frequency current noise within the Hartree approximation, which is an extension of the previous results obtained by Hershfield on zero frequency shot noise.
\end{abstract}
 \maketitle

\newpage

\section{introduction}

The transport properties of mesoscopic conductors have attracted wide research interest, and are of great importance for future nanoscale electronic applications. One usually probes the dynamics of electron in the conductors by measuring the dc and ac conductance in the linear response regime or driving the systems to the nonlinear out-of equilibrium case. For instance, Gabelli et al.\cite{Gabelli} measured the ac conductance of a mesoscopic RC circuit at GHz frequency, and found the violation of Kirchhoff's law of impedance addition.  Recent experimental advance will enable to probe the electronic processes in these systems in the high frequency region reaching the intrinsic time scales of the electron dynamics, therefore new interesting physical properties are expected to be observed.\cite{Chevallier}

Some important progresses in this field are based on a wisdom that more information about electron dynamics can be obtained by measuring the current fluctuations or the higher current moments in these systems. \cite{Landauer,Blanter}  The theory of  full counting statistics (FCS) of electron current, which describes the probability distribution of transmitted charge during a fixed time interval through a mesoscopic conductor,  were developed within the scattering formulation,\cite{Levitov1993,Levitov1996,Nazarov} and the Hamiltonian formulation for FCS were constructed based on the Anderson impurity model.\cite{Gogolin,Schmidt,Levitov2004}   It was also shown that the real part of ac conductance is related to the asymmetric parts of the frequency dependent current noise  by a non-equilibrium fluctuation-dissipation theorem. \cite{Safi,Billangeon}  Therefore, at present there are intensive research actives on the study of FCS problems both in quantum dot systems \cite{Bagrets,Utsumi2007,Kambly} and diffusive conductors. \cite{Pilgram,Lee}

For the study of the time-dependent and finite frequency transport properties of mesoscopic systems, Blanter and B\"uttiker \cite{Blanter} emphasized the importance of considering electron interaction effects, and pointed out that the current measured in frequency dependent transport experiment is a sum of the particle current and displacement current, while the displacement current is zero in the static case, but is essential for finite frequency transport. A theoretical approach that can address the finite frequency electron transport and take into account the various electron correlation effects as well as the current conservation condition explicitly is highly worthy of investigation.\cite{Pedersen,Wei}

A prototype mesoscopic system of nonequilibrium electron transport with strong Coulomb interaction is a single quantum dot coupled to left and right leads, which can be described by the Anderson impurity model. The current through the quantum dot\cite{H1992a} had been calculated by performing a second order perturbation theory of the Coulomb interaction strength $U$. However, the authors\cite{H1992a} observed that the current is not conserved in this approximation when the dot level is tuned away from the particle-hole symmetry point, and pointed out the necessity of treating interaction terms by a current conservation approximation.  Hershfield \cite{H1992b} studied the current fluctuations in the Anderson impurity model subsequently, and obtained a formula for interaction effect on zero frequency shot noise in the Hartree approximation.  During the last decade, the shot noise and the FCS of current of the Anderson impurity model in the low-temperature Kondo regime have attracted a great deal of research efforts and interests, but we will restrict our consideration only to the resonant tunnelling regime in this paper. For a noninteracting resonant tunneling model, exact analytical formula of the finite frequency noise spectrum had been obtained.\cite{Chen}  The nonsymmetrized noise spectrum of the resonant tunneling model had been studied recently within the scattering formulation \cite{Entin-Wohlman,Rothstein}, where some step and dip structures at finite frequency were shown, and the Hartree-Fock theory was applied to multi-level quantum dot system.\cite{Gabdank}

In the present work, we have formulated the theory of electron transport through quantum dot based on the nonequilibrium self-consistent perturbation method,\cite{Baym}  which can guarantee the gauge invariance and current conservation condition for arbitrary time-dependent  external potential. We have studied the effect of the potential fluctuations in the quantum dot on the finite frequency noise within the Hartree approximation, and obtained an analytical equation for the interaction effect on the current noise, which is an extension of Hershfield's result \cite{H1992b} to the finite frequency case.  We will show that various correlation functions of nonequilibrium system including the vertex correction terms can be calculated in a systematic way by using functional derivations defined on the closed time contour and the external counting field method. Actually, the functional approach has been applied to a wide range of problems in mesosocopic system in the literatures, e.g.  current and noise characteristics for single-electron transistors in the Coulomb blockade regime \cite{Utsumi2003,Oh} or the Kondo regime,\cite{Moca}  photoassisted shot noise in mesoscopic conductor, \cite{Chevallier} charge transport to chaotic cavity\cite{Macedo} and the time evolution of non-equilibrium quantum dot system,\cite{Sexty} etc.

This paper is organized as follows: In section II, we discuss the current conservation condition and the generalized Ward identity for quantum dot based on the Anderson impurity model.  In section III, the formula for finite frequency noise spectrum including the interaction correction term in the Hartree approximation is obtained.  Section IV is devoted to the numerical calculations. In section V we summarize the results of this work.

\section {Self-consistent perturbation theory and the generalized Ward identity}
  We consider the electrons transport through a single level quantum dot in the presence of external ac fields, the system will be described by the following Anderson impurity model\cite{Ng,Ding}
\bn
H&=&\sum_{k\eta\sigma}\epsilon_{k\eta}(t)c^\dagger_{k\eta\sigma}c_{k\eta\sigma}+
\sum_\sigma \epsilon_d(t) d^\dagger_\sigma d_\sigma + Un_{d\uparrow}n_{d\downarrow} \nonumber\\
&& +\sum_{k\eta\sigma}\left [ t_\eta e^{i \lambda_\eta
(t)}c^\dagger_{k\eta\sigma}d_\sigma+H.c.\right ]\;,
\en
where $\eta=L,R$ denotes the left and right leads, $\epsilon_{k\eta}(t)=\epsilon_{k\eta}+v_\eta(t)$ and $\epsilon_d(t)=\epsilon_d+v_0(t)$, with $v_\eta(t)$ and $v_0(t)$ being the ac potentials in leads and in the dot, respectively. $\lambda_\eta(t)$ is the gauge potential coupled to the tunneling current from the lead $\eta$ to the dot. The strong Coulomb interaction term in the Hamiltonian prevent this model from an exact solution. But we can approach this problem by doing perturbation theory on the Schwinger-Keldysh closed time path contour. Using the nonequilibrium Dyson equation, we write the equation of motion for the nonequilibrium Green's function of the quantum dot as follows
\bq
 [ i{\frac {\partial}{\partial
t}}-\epsilon_d(t) ]G_{d\sigma}(t,t')=\delta(t,t')+\int dt_1
\Sigma(t,t_1) G_{d\sigma}(t_1, t')\;.
\eq
It should be emphasized that the time variables $t$ and $t'$ can be either on the forward or the backward branch of the time contour, and the integration over $t_1$ is also defined on the closed time path contour. The self-energy $\Sigma(t,t')$ is obtained in the framework of self-consistent perturbation theory, and can be divided into two terms
\bq
\Sigma(t,t')=\Sigma^{(0)}(t,t')+\Sigma_{U}(t,t') \;,
\eq
where $\Sigma^{(0)}(t,t')$ is the dot level self-energy contributed from the tunneling between the leads and the quantum dot
\bq
\Sigma^{(0)}(t,t')=\sum_{k\eta}|t_\eta|^2
\bar{g}_{k\eta}(t,t')e^{-i[\lambda_\eta(t)-\lambda_\eta(t')+\int^t_{t'}dt_1
v_\eta(t_1)]}\;,
\eq
with $\bar {g}_{k\eta}(t,t')$ being the bare Green's function of the lead without external ac potential field. $\Sigma_U(t,t')$ is the self-energy due to Coulomb interaction. In the self-consistent perturbation theory, it is a functional of the full Green's functions of the quantum dot. In order to illustrate the method of calculation, we consider only the first order approximation, and the interaction self-energy is given by the Hartree term
\bq
\Sigma_U(t,t')=U\langle n_{d\bar\sigma}(t)\rangle\delta(t,t')\;.
\eq

We first study the gauge transformation properties and current conservation condition of the Green's functions of quantum dot. By making a transformation\cite{Ng} $G_{d\sigma}(t,t')=\bar G_{d\sigma}(t,t')e^{-i\int_{t}^{t'} dt_1 v_0(t_1)} $,  the equation of motion for $\bar G_{d\sigma}(t,t')$ will be given by
\begin{equation}
[i{\frac {\partial} {\partial t}}-\epsilon_d]\bar
G_{d\sigma}(t,t')=\delta(t,t')+\int dt_1 \bar\Sigma(t,t_1) \bar
G_{d\sigma}(t_1, t')\;,
\end{equation}
where the self-energy $\bar\Sigma(t,t_1)=\bar\Sigma^{(0)}(t,t')+\bar\Sigma_U(t,t')$ with
\bq
\bar\Sigma^{(0)}(t,t')=\sum_{k\eta}|t_\eta|^2
\bar{g}_{k\eta}(t,t')e^{-i\phi_\eta(t,t')}\;,
\eq
and in the Hartree approximation
\bq
\bar\Sigma_U(t,t')=U \langle\bar n_{d\bar\sigma}(t)\rangle\delta(t,t')\;,
\eq
where the phase factor
$\phi_\eta(t,t')=\lambda_\eta(t)-\lambda_\eta(t')+\int^t_{t'}dt_1
[v_\eta(t_1)-v_0(t_1)]$, and  $\bar G_{d\sigma}(t,t')$ is a gauge transformation invariant quantity.  If one consider a gauge transformation: $v_0(t)\rightarrow
v_0(t)+\partial_t \tilde\Lambda(t)$, $\lambda_\eta(t)\rightarrow \lambda_\eta(t) +\tilde\Lambda(t)$.
 then it is easy to see that the phase factor $\phi_\eta(t,t')$, the self-energy $\bar\Sigma(t,t')$ and
$\bar G_{d\sigma}(t,t')$ are all gauge transformation invariant.
Therefore, the Green's function $G_{d\sigma}(t,t')$ transforms as
\bq
G_{d\sigma}(t,t';\tilde\Lambda)=e^{-i\tilde\Lambda(t)}G_{d\sigma}(t,t'
)e^{i\tilde\Lambda(t')}\;.
\eq
The above gauge transformation is directly related to the current conservation condition in the quantum dot. Since under this gauge transformation, the change of the Hamiltonian to the first order of $\tilde\Lambda$ is given by
\bq
\delta H(t)= n_d(t)\partial_t\tilde\Lambda(t)+\sum_\eta
j_\eta(t)\tilde\Lambda(t)\;,
\eq
where $n_d(t)$ and $j_\eta(t)$ are the operators of the charge number in the dot and the tunneling current from the lead $\eta$ to the dot, respectively. The gauge transformation invariance of the action leads to the continuity equation
\bq
\partial_t \langle n_d(t)\rangle-\sum_\eta \langle j_\eta(t)\rangle=0\;.
\eq
In the out-of equilibrium steady state, the occupation number $\langle n_d(t)\rangle$ is time independent, and the current conservation condition $\sum_\eta \langle j_\eta\rangle=0$ is satisfied.

Next we will follow the procedure as given in the Ref.\cite{Leeuwen} to give a derivation of the generalized Ward identity\cite{Ward} for this quantum dot system, which is closely related to the current conservation condition. In the spin degenerate case, the spin index $\sigma$ will be omitted.  We consider the changes in the Green's function induced by the  gauge transformation. From Eq.(9) the first-order change in $G$ is
\bq
\delta G(t,t')=-i\left [\tilde\Lambda(t)-\tilde\Lambda(t')\right ]G(t,t')\;,
\eq
and it leads to the equation
\bn
\int dt_1 \left [{\frac {\delta G(t,t')} {\delta
v_0(t_1)}}\partial_{t_1}\tilde\Lambda(t_1)+\sum_\eta {\frac {\delta
G(t,t')}{\delta \lambda_\eta(t_1)}}\tilde\Lambda(t_1)\right ]
\nonumber\\
=-i\left [\tilde\Lambda(t)-\tilde\Lambda(t')\right ]G(t,t')\;,
\en
where the functional derivatives of $G$ can be denoted as
$\Lambda_0$ and $\Lambda_\eta$, and they correspond to the time-ordered operator products as follows
\bq
\Lambda_0(t,t';t_1)={\frac {\delta G(t,t')} {\delta
{v_0(t_1)}}}=-\langle T_C[d_\sigma(t)d^\dagger_\sigma(t')n_d(t_1)]\rangle\;,
\eq
\bq
\Lambda_\eta(t,t';t_1)={\frac {\delta G(t,t')} {\delta
{\lambda_\eta(t_1)}}}=-\langle T_C[d_\sigma(t)d^\dagger_\sigma(t')j_\eta(t_1)]\rangle\;.
\eq
An integration by parts in Eq. (13), and demanding that the equation is satisfied for arbitrary $\tilde\Lambda$, straightforwardly leads to the well-known generalized Ward identity
\bn
\partial_{t_1}\Lambda_0(t,t';t_1)&-&\sum_\eta
\Lambda_\eta(t,t';t_1)\nonumber\\
&=&i[\delta(t,t_1)-\delta(t',t_1)]G(t,t')\;.
\en
This identity leads to a relation between the vertex functions and the self-energy, which can be demonstrated explicitly by introducing the following vertex functions
\bq
\Gamma_0(t,t';t_1)=-{\frac {\delta G^{-1}(t,t')}{\delta v_0(t_1)}}\;,
\eq
\bq
\Gamma_\eta(t,t';t_1)=-{\frac {\delta G^{-1}(t,t')} {\delta
\lambda_\eta(t_1)}}\;.
\eq
One can see that these vertex functions are related to the time-ordered operators
\bq
\Lambda_0(t,t';t_1)=\int dt_2 dt_3
G(t,t_2)\Gamma_0(t_2,t_3;t_1)G(t_3,t')\;,
\eq
\bq
\Lambda_\eta(t,t';t_1)=\int dt_2 dt_3
G(t,t_2)\Gamma_\eta(t_2,t_3;t_1)G(t_3,t')\;.
\eq
Thereby, the generalized Ward identity Eq. (16) can be rewritten in terms of the vertex functions
\bn
\partial_{t_1}\Gamma_0(t,t';t_1)&-&\sum_\eta
\Gamma_\eta(t,t';t_1)\nonumber\\
&=&i[\delta(t',t_1)-\delta(t,t_1)]G^{-1}(t,t')\;.
\en
This equation relates the vertex functions to the self-energy, since the inverse of Green's function $G^{-1}=G_0^{-1}-\Sigma$ is given explicitly as
\bq
G^{-1}(t,t')=[i\partial_t-\epsilon_d-v_0(t)]\delta(t,t')-\Sigma^{(0)}(t,t')-\Sigma_U(t,t')\;.
\eq
The generalized Ward identity implies the gauge invariance and current conservation condition in this problem. Therefore it should be satisfied when we are investigating the current fluctuations or time-dependent electron transport properties in this system by making approximation calculation of the self-energy or the vertex functions.

\section{ Current fluctuations and the Hartree approximation}

 In this section,  we study the current fluctuation and statistical problems in this quantum dot system. It is well known that the central quantity in FCS calculations \cite{Levitov1993,Levitov1996} is the cumulant generating function $\chi ({\bf \lambda})=\sum_{\bf Q}e^{i{\bf Q \bf \lambda}}P({\bf
Q})$, where ${\bf\lambda}=(\lambda_1,
\ldots,\lambda_N)$ are the counting fields and $P({\bf Q})$ is the probability for the charge ${\bf Q}=(Q_1,\ldots, Q_N)$ to be transferred through the respective channel\cite{Schmidt} during the measuring time $\it T $. For noninteracting electron system, the generating function is given by the Levitov-Lesovik formula\cite{Levitov1993} within scattering matrix approach. It is observed that the cumulant generating function can be generalized to the system with time-dependent counting fields, and it can be written in terms of the nonequilibrium Green's function defined on the closed time path contour as follows
\begin{equation}
\ln \chi({\bf \lambda})=\mathrm{Tr}\ln G^{-1} -\mathrm{Tr} \Sigma_U G +\Phi(G)\;,
\end{equation}
where $G^{-1}$ is the inverse of the full Green's function of the quantum dot given explicitly by Eq.(22).   $\Phi$ a functional potential constructed by summing over irreducible self-energy diagrams closed with an additional Green function line.\cite{Baym} The interaction self-energy  $\Sigma_U$ can be obtained from the functional $\Phi$ by
\begin{equation}
\Sigma_U(t,t')={\frac {\delta\Phi} {\delta G(t',t)}}\;.
\end{equation}
One can verify that when the counting fields $\lambda$ are assumed to be time-independent and the system is in the noninteracting case, the above generating functional arrives at the Levitov-Lesovik formula.

The electron current tunneling from the lead $\eta$ to the quantum dot can be obtained by a functional derivative of $\chi (\lambda)$ with respect to $\lambda_\eta (t)$
\begin{eqnarray}
&&\langle I_\eta(t)\rangle=i{\frac e \hbar}{\frac {\delta \ln \chi(\lambda)} {
{\delta\lambda_\eta(t)}} }\nonumber\\
&&=-i{\frac e  \hbar}\int dt_1 dt_2
G(t_1,t_2)\Gamma^{(0)}_\eta(t_2,t_1;t)\;.
\end{eqnarray}
Here the bare current vertex function $\Gamma^{(0)}_\eta(t_2,t_1;t)$ is given by
\begin{eqnarray}
\Gamma^{(0)}_\eta(t_2,t_1;t)&=&{\frac {\delta\Sigma^{(0)}(t_2,t_1)} {\delta\lambda_\eta(t)}}
\nonumber\\
&=&i\left [\delta(t_1,t)-\delta(t_2,t)\right ]\Sigma^{(0)}_\eta(t_2,t_1)\;.
\end{eqnarray}
Therefore the current through quantum dot is given as\cite{Oh,Macedo}
\begin{equation}
\langle I_\eta(t)\rangle={\frac e  \hbar}\int dt_1 [G(t,t_1)\Sigma^{(0)}_\eta(t_1,t)
-\Sigma^{(0)}_\eta(t,t_1)G(t_1,t) ]\;.
\end{equation}
By using the operational rules given by Langreth for contour integration, \cite{Langreth} it is not difficult to prove that this formula is exactly equivalent to the current formula obtained by Jauho et al.\cite{Jauho} for the time-dependent electron transport through an interacting quantum dot.

We can introduce the interaction induced current vertex function which is related to the self-energy of Coulomb interaction
\begin{equation}
\Gamma^U_\eta(t_2,t_1;t)={\frac {\delta\Sigma_{U}(t_2,t_1)} {\delta\lambda_\eta(t)}}\;.
\end{equation}
Then the vertex function defined in Eq.(18) is given by
\begin{equation}
\Gamma_\eta(t_2,t_1;t)=\Gamma^{(0)}_\eta(t_2,t_1;t)+\Gamma^U_\eta(t_2,t_1;t)\;.
\end{equation}
The  current formula Eq. (25) indicates that the  current is contributed solely from the bare current vertex. There is no contribution of interaction vertex correction to the current in this self-consistent perturbation approach. However, we will show in the following that the interaction vertex indeed influences the current fluctuations.

 Next, we calculate the current-current correlation functions on the time contour and find that they can be represented as the sum of two terms
\bn
D_{\eta\eta'}(t,t')&\equiv&  \langle T_C\delta I_\eta(t)\delta I_{\eta'}(t')\rangle
=-{\frac {e^2} \hbar}{\frac {\delta^2
\ln \chi(\lambda)}
{\delta\lambda_\eta(t)\delta\lambda_\eta(t')}}
\nonumber\\
&=&D^{(0)}_{\eta\eta'}(t,t')+ D^{(c)}_{\eta\eta'}(t,t')\;,
\en
where the bare term is
\begin{widetext}
\bn
D^{(0)}_{\eta\eta'}(t,t')&=&{\frac {e^2} \hbar}\delta_{\eta\eta'}\left [G(t,t')\Sigma^{(0)}_\eta(t',t)
+\Sigma^{(0)}_\eta(t,t')G(t',t)\right ]
\nonumber\\
&&+{\frac {e^2} \hbar}\int dt_1 dt_2 dt_3 dt_4
\left [G(t_1,t_2)\Gamma^{(0)}_{\eta'}(t_2,t_3;t')G(t_3,t_4)\Gamma_{\eta}^{(0)}(t_4,t_1;t)\right ]\;,
\en
and the interaction induced vertex correction term to the current correlation is given by
\bq
D^{(c)}_{\eta\eta'}(t,t')={\frac {e^2} \hbar}\int dt_1 dt_2 dt_3 dt_4
\left [G(t_1,t_2)\Gamma_{\eta'}^U(t_2,t_3;t')G(t_3,t_4)\Gamma_{\eta}^{(0)}(t_4,t_1;t)\right ]\;.
\eq

Among the various current correlation functions, the correlation function for current noise is of particular interest, since the frequency dependent noise spectrum of current contains the intrinsic dynamics information of this quantum dot system. In a steady state without external time-dependent potential, the symmetrized noise spectrum $S_{\eta\eta'}(\omega)$ is given by the Fourier transform of the correlation function of current operators  $S_{\eta\eta'}(t,t')=\langle \delta I_\eta(t)\delta
I_{\eta'}(t')\rangle+\langle \delta I_{\eta'}(t')\delta I_{\eta}(t)\rangle$. It is noted that the correlation function for current noise can be written as
\bq
S_{\eta\eta'}(t,t') = D^{>}_{\eta\eta'}(t,t')+D^{<}_{\eta\eta'}(t,t')
= S^{(0)}_{\eta\eta'}(t,t')+ S^{(c)}_{\eta\eta'}(t,t')\;,
\eq
where $S^{(0)}_{\eta\eta'}(t,t')$ and  $S^{(c)}_{\eta\eta'}(t,t')$ are contributed from the bare term and the interaction induced vertex correction term, respectively.

The bare term $S^{(0)}_{\eta\eta'}(t,t')$ is obtained straightforwardly by using Langreth's analytical continuation rules.\cite{Langreth}  In the absence of external ac potential, we can transform it to the frequency space, and  express it in terms of the Green's functions of quantum dot explicitly as \cite{Dong,Lopez}
\bn
S^{(0)}_{\eta\eta'}(\omega)&=&{\frac {e^2} \hbar}\int {\frac {d\omega_1} {2\pi}}\bigg \{ \delta_{\eta\eta'}
i\Gamma_\eta \left [ n_\eta(\omega_1)G^>(\omega_1+\omega)-(1-n_\eta(\omega_1+\omega))G^<(\omega_1)\right ]
\nonumber\\
&&-\Gamma_\eta\Gamma_{\eta'} \Big \{ n_\eta(\omega_1)(1-n_{\eta'}(\omega_1+\omega) )G^r(\omega_1)G^r(\omega_1+\omega)+
n_{\eta'}(\omega_1) (1-n_{\eta}(\omega_1+\omega))G^a(\omega_1)G^a(\omega_1+\omega)
\nonumber\\
&&+  [ n_{\eta'}(\omega_1)G^a(\omega_1)-n_\eta(\omega_1)G^r(\omega_1) ]G^>(\omega_1+\omega)
+G^<(\omega_1) [(1-n_{\eta'}(\omega_1+\omega))G^r(\omega_1+\omega)\nonumber\\
&&-(1-n_{\eta}(\omega_1+\omega))G^a(\omega_1+\omega)]
-G^<(\omega_1)G^>(\omega_1+\omega)\Big \} \bigg \} +\bigg \{\omega\rightarrow -\omega  \bigg \}\;.
\en

In order to obtain the interaction effect on the noise spectra, we have to calculate the vertex function by functional derivation of the interaction self-energy with respect to the counting field: $ \Gamma^U_\eta(t_1,t_2;t)=  {\frac {\delta\Sigma_{U}(t_1,t_2)} {\delta\lambda_\eta(t)}}  $, where the interaction self energy $\Sigma_{U}(t_1,t_2)$ is given by Eq.(5) in the Hartree approximation. The technical details of our calculation is presented in Appendix B. After calculating the vertex function and transform it to the frequency space, we can obtain the interaction correction to the finite frequency current correlation function $S^{(c)}_{\eta\eta'}(\omega)$ as follows
\bn
S^{(c)}_{\eta\eta'}(\omega)&=&{\frac {e^2} \hbar}\bigg [\chi^{r,(0)}_{j_{\eta}n}(\omega){\frac U  {1-U\chi^{r,(0)}_{nn}(\omega)}}S^{(0)}_{nj_{\eta'}}(\omega)+S^{(0)}_{j_\eta n}(\omega) {\frac U  {1-U\chi^{a,(0)}_{nn}(\omega)}}\chi^{a,(0)}_{n j_{\eta'}}(\omega)
\nonumber\\
&&+\chi^{r,(0)}_{j_\eta n}(\omega) {\frac U  {1-U\chi^{r,(0)}_{nn}(\omega)}}S^{(0)}_{nn}(\omega)
{\frac U  {1-U\chi^{a,(0)}_{nn}(\omega)}}\chi^{a,(0)}_{n j_{\eta'}}(\omega)\bigg ]\;,
\en
\end{widetext}
where the various correlation and response functions are given in Appendix A. This equation is the central result of our paper. It is a generalization of the zero frequency noise result obtained by Hershfield\cite{H1992b} to the finite frequency case, and can be interpreted as the current noise contributed from the coupling of  density fluctuations in the quantum dot to the current fluctuations. The first term in the Eq.(35) indicates that the correlation between the density and the current $S^{(0)}_{nj_\eta'}(\omega)$ can propagate forward in time via $\chi^{r,(0)}_{nn}(\omega) $ and $\chi^{r,(0)}_{j_{\eta}n}(\omega)$ to produce current fluctuations. The second term represents the correlation $S^{(0)}_{j_{\eta}n}(\omega)$ propagates backward in time on the backward branch to produce current fluctuations. In the last term the correlation between the densities at the quantum dot $S^{(0)}_{nn}(\omega)$ propagates forward and backward on the closed time contour simultaneously and gives rise to current fluctuations.

\begin{figure}[hbtp]
\includegraphics[width=\columnwidth, height=7cm, angle=0]{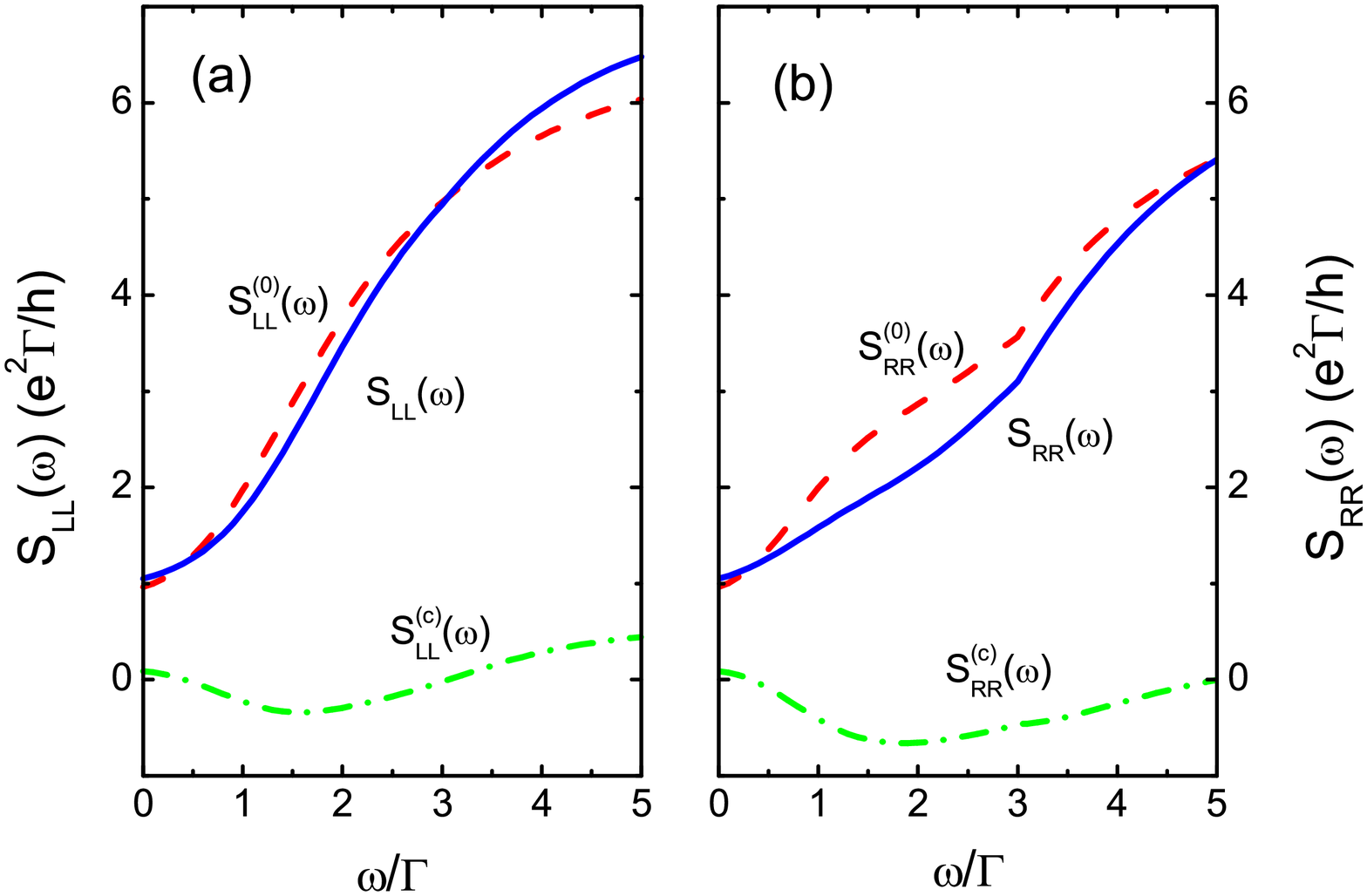}
\caption{ (Color online) The  noise spectra for quantum dot system in the symmetric coupling case. (a) for left lead, we plot the bare noise spectrum $S^{(0)}_{LL}(\omega)$ (dashed line), the interaction correction term $S^{(c)}_{LL}(\omega)$ (dash-dotted line), and the noise spectrum after correction $S_{LL}(\omega)$ (solid line); (b) the same for the right lead. Here, we take the parameters in the Anderson impurity model as $\epsilon_d=-1.0$,  $U=4.0$ in units of the coupling strength $\Gamma$, and we assume $\Gamma_L=\Gamma_R=1.0$ for the symmetric case.}
\end{figure}

\section{numerical results}
 To get better understanding of the interaction effect on the current noise spectrum for this quantum dot system, we will present some numerical calculations of the current noise at zero temperature.  In our calculation, we take the coupling strength between the leads and quantum dot $\Gamma$ as the units of the energy, and take the parameters $\epsilon_d=-1.0$, $U=4.0$ and the bandwidth $D=100$. The applied bias voltage $\Delta\mu=3.0$, with $\mu_L=-\mu_R=\Delta\mu/2$.  For the system with symmetric coupling  strength $\Gamma_L=\Gamma_R=\Gamma$, we plot the current noise spectra for left and right leads in Fig.1(a) and (b), respectively. Fig.1 (a) shows that the bare noise spectrum $S^{(0)}_{LL}(\omega)$ is positive and is an increasing function of the frequency. The interaction correction term $S^{(c)}_{LL}(\omega)$ can have negative values at an intermediate finite frequencies region. It is observed that the interaction correction for shot noise at zero frequency has a rather small positive value, which agrees with the previous result,\cite{H1992b}  but the interaction correction becomes negative and more significant when the frequency increases, and it goes to positive value again in the large frequency region. The maximum influence of interaction correction is obtained at the frequency which is largely determined by the energy difference between the renormalized dot level $\tilde\epsilon_d=\epsilon_d+U \langle n_{d\bar\sigma}\rangle $ and the Fermi level of the leads.  The sum of $S^{(0)}_{LL}(\omega)$ and $S^{(c)}_{LL}(\omega)$ gives the noise spectrum after interaction correction, which is a monotonously increasing function of the frequency. Fig.1 (b) shows the noise spectra for the right lead (the drain side of this system). These noise spectra have more prominent features than that of the left lead. One can find a significant dip for the total noise spectrum $S_{RR}(\omega)$ at the frequency equal to the applied bias voltage ($\omega=\Delta \mu$=3.0).

 The various bare correlation and response functions utilized in the calculation of interaction effect of noise spectra are plotted in Fig.2. As shown in Fig.2 (a) that the density fluctuation spectrum $S^{(0)}_{nn}(\omega)$ of the quantum dot always has real positive values at finite frequencies.  The real part of density response function $\chi^{(0)}_{nn}(\omega)$  is negative at low frequency, which indicates that the screening effect of electron  decreases the Coulomb interaction strength $U$ on the quantum dot. The imaginary part of  $\chi^{(0)}_{nn}(\omega)$ remains negative for all frequencies due to the analytical properties of the density-density response function. The correlation and response functions between the density operator on the dot and the current operator for the left and right lead are plotted in Fig.2(b) and (c), respectively.

\begin{figure}[hbtp]
\includegraphics[width=\columnwidth, height=7cm, angle=0]{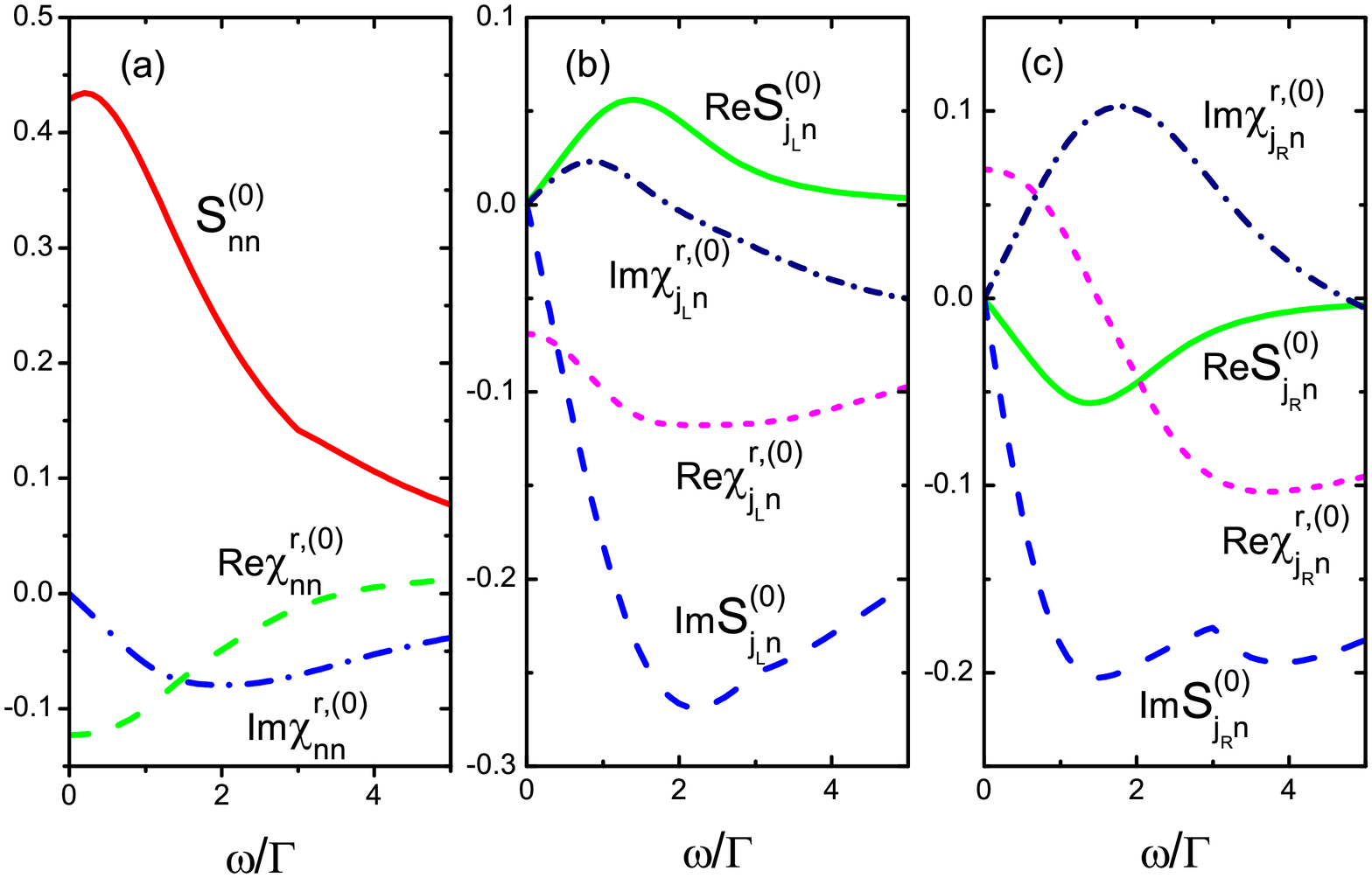}
\caption{(Color online) The  spectra of various correlation and response functions involved in the calculation of the interaction effect on noise spectra in the symmetric coupling case (in units of $e^2/\Gamma$). (a) The density correlation function $S^{(0)}_{nn}(\omega)$, the real and imaginary parts of the density response function $\chi^{(0)}_{nn}(\omega)$ in the quantum dot. (b) The real and imaginary parts of the correlation function $S^{(0)}_{j_L n}(\omega)$  and response function $\chi^{r,(0)}_{j_L n}(\omega)$ for the left lead. (c) The correlation function and response function of the right lead.
The parameters used in the calculation are the same as in Fig.1. }
\end{figure}

 Fig.3 shows the current noise spectra for asymmetrically coupled quantum dot system. Since the coupling strength $\Gamma_L\gg\Gamma_R $, we find that the magnitude of the current fluctuations in the right lead plotted in Fig.3
(b) is much less than that of the left lead in Fig.3(a), because of the tunneling rate between the right lead and quantum dot is much less than that of the left lead. The interaction correction terms also have negative value regions at finite frequencies both for the left and right leads. The noise spectrum in the drain lead (right lead) shows an evident dip structure at the frequency equal to the bias voltage. One can expect this kind of prominent features of noise spectrum can be detected in experiments.

\begin{figure}[hbtp]
\includegraphics[width=\columnwidth,height=7cm,angle=0]{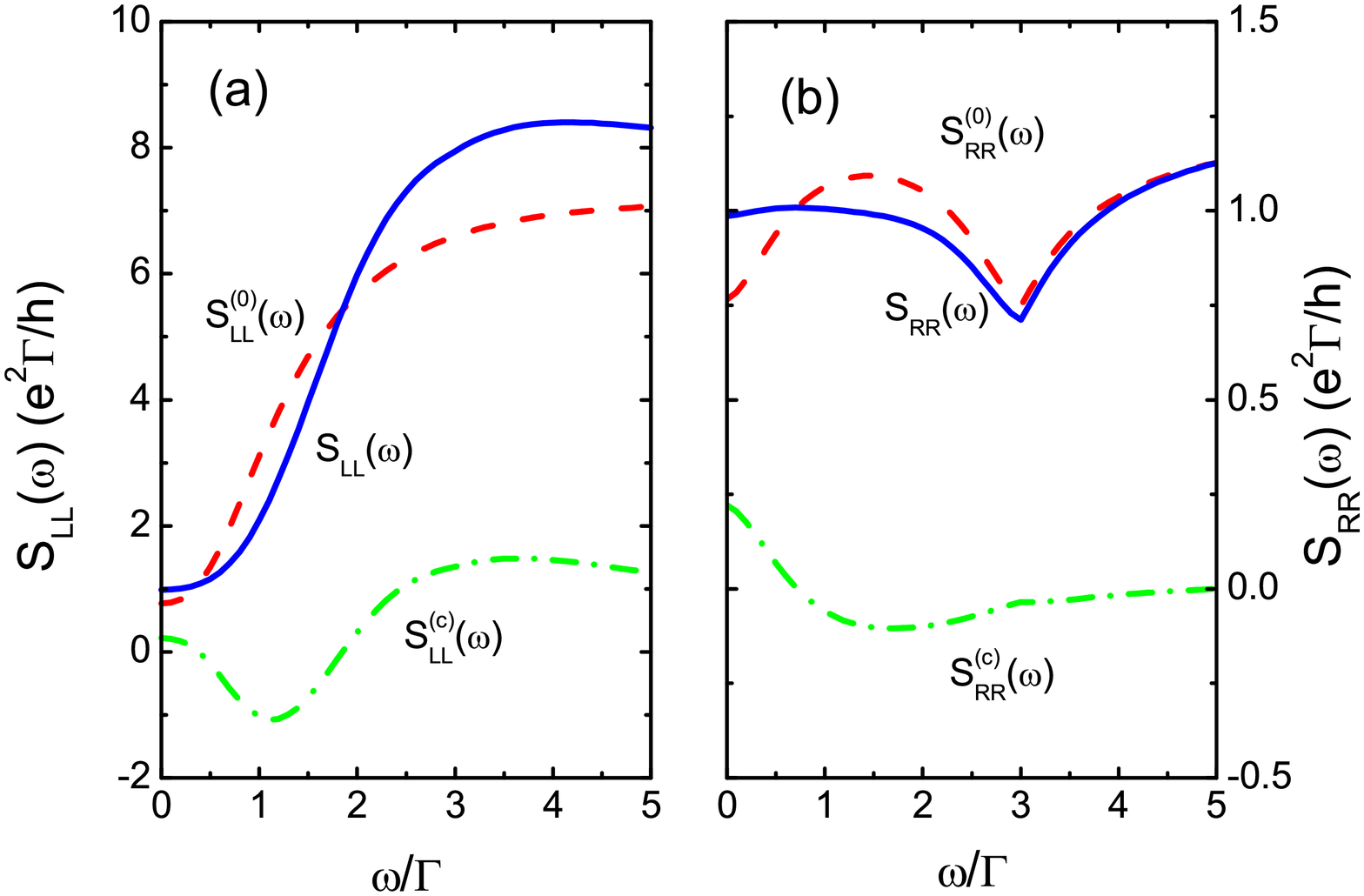}
\caption{ (Color online) The  noise spectra for quantum dot system in the asymmetric coupling case . (a) For the left lead, we plot the bare noise spectrum $S^{(0)}_{LL}(\omega)$ (dashed line), the interaction correction term $S^{(c)}_{LL}(\omega)$ (dash-dotted line), and the noise spectrum after correction $S_{LL}(\omega)$ (solid line); (b) The same for the right lead.  We take the parameters in the Anderson model as $\epsilon_d=-1.0$,  $U=4.0$ in units of the coupling strength $\Gamma$, and assume $\Gamma_L=1.0, \Gamma_R=0.2$ for the asymmetric case. }
\end{figure}

\section{conclusions}
In this work, we have investigated the problem of electron transport through quantum dot in the framework of nonequilibrium self-consistent perturbation theory and examined the current conservation condition. Based on the Anderson impurity model, we gives the current and current fluctuations formulae, which are valid in the presence of arbitrary external time-dependent potentials, by using nonequilibrium generating functional and the functional derivation method. We have calculated the interaction effect on the finite frequency noise spectrum of Anderson impurity model by taking into account the interaction vertex correction term with the Hartree approximation, and obtained an analytical equation for the noise correction term at finite frequency, which corresponds to a generalization of the previous result on zero frequency shot noise.  We have focused our attention on the symmetrized noise spectrum, one can expect that the nonsymmetrized noise spectrum and the ac conductance can also be studied within the formulation presented in this paper. We believe that the self-consistent perturbation theory on the Schwinger-Keldysh contour can lead to a unified approach to many interesting problems in  nonequilibrium electron transport through mesoscopic systems, and will give us deep understanding of the current fluctuation and energy dissipation phenomena.\cite{Das} The functional method provides a convenient way to study the statistics of current fluctuations. We note that the Hartree approximation can only account for some interaction effects on the electron current noise in the resonant tunnelling regime.  It is expected that future research work can treat the interacting effect beyond the Hartree approximation, and give us more information about the interaction effect on the out-of equilibrium dynamics of electrons in the Coulomb blockade regime as well as the low-temperature Kondo regime.

\begin{acknowledgments}
This work was supported by Projects of the National Basic Research Program of China (973 Program) under Grant No. 2011CB925603, and the National Science Foundation of China, Specialized Research Fund for the Doctoral Program of Higher Education (SRFDP) of China.
\end{acknowledgments}

\appendix
\begin{appendix}
\section{}
The response and correlation functions of density operator are given by
\bq
\chi^{r}_{nn}(t,t')=-i\theta(t,t')\langle[n_{d\sigma}(t), n_{d\sigma}(t')]\rangle\;,
\label{A1}
\eq
and
\bq
S_{nn}(t,t')=\langle\{\delta n_{d\sigma}(t), \delta n_{d\sigma}(t')\}\rangle\;,
\eq
respectively. The current-density response function and correlation function
\bq
\chi^{r}_{j_\eta n}(t,t')=-i\theta(t,t')\langle[j_{\eta\sigma}(t), n_{d\sigma}(t')]\rangle\;,
\eq
and
\bq
S_{j_\eta n}(t,t')=\langle\{\delta j_{\eta\sigma}(t), \delta n_{d\sigma}(t')\}\rangle\;.
\eq
By neglecting the vertex correction term induced by interaction effect,
we can write the above functions in the frequency space explicitly as follows
\begin{widetext}
\bq
\chi^{r,(0)}_{nn}(\omega)=\int {\frac {d\omega_1} {2\pi}}(-i) [G^a(\omega_1)G^{<}(\omega_1+\omega)+G^<(\omega_1)G^r(\omega_1+\omega)]\;,\\
\eq
\bq
S^{(0)}_{nn}(\omega)=\int {\frac {d\omega_1} {2\pi}} [G^<(\omega_1)G^>(\omega_1+\omega)+G^>(\omega_1)G^<(\omega_1+\omega)]\;.
\eq

\bq
\chi^{r,(0)}_{j_\eta n}(\omega)=\int {\frac {d\omega_1} {2\pi}} \left \{   i \Gamma_\eta n_\eta(\omega_1)[G^r(\omega_1)G^r(\omega_1+\omega)
    -G^a(\omega_1)G^a(\omega_1-\omega)]+ i \Gamma_\eta G^<(\omega_1)[G^r(\omega_1+\omega)+G^a(\omega_1-\omega)] \right \}\;,
\eq

\bn
S^{(0)}_{j_\eta n}(\omega)&=&\int {\frac {d\omega_1} {2\pi}} \Gamma_\eta \{ (1-n_\eta(\omega_1))G^r(\omega_1)
G^<(\omega_1+\omega)-n_\eta(\omega_1)G^r(\omega_1)G^>(\omega_1+\omega)
\nonumber\\
&+& n_\eta(\omega_1+\omega)G^>(\omega_1)G^a(\omega_1+\omega)-(1-n_\eta(\omega_1+\omega))G^<(\omega_1)G^a(\omega_1+\omega)
\nonumber\\
&-&G^<(\omega_1)G^>(\omega_1+\omega)- G^>(\omega_1)G^<(\omega_1+\omega)\}\;.
\en

In the Hartree approximation, the retarded/advanced Green's function $G^{r/a}(\omega_1)=
1/[ \omega_1-\epsilon_d-U\langle n_{d\bar\sigma}\rangle\pm i\sum_\eta\Gamma_\eta /2] $. The other Green's functions $G^<(\omega_1)=G^r(\omega_1)[i\sum_\eta \Gamma_\eta n_\eta(\omega_1) ]G^a(\omega_1)$, and $G^>(\omega_1)=G^r(\omega_1)[-i\sum_\eta \Gamma_\eta (1-n_\eta(\omega_1)) ]G^a(\omega_1)$.

\section{}
The interaction vertex function $ \Gamma^U_\eta(t_1,t_2;t)=  {\frac {\delta\Sigma_{U}(t_1,t_2)} {\delta\lambda_\eta(t)}}  $, and
the self-energy in the Hartree approximation is given by
\bq
\Sigma_{U}(t_1,t_2)=U\langle n_{d\bar\sigma}(t_1)\rangle\delta(t_1,t_2)=-iU G_{\bar\sigma}(t_1,t_1^+)\delta(t_1,t_2)\;.
\eq
In the spin degenerate case,  we will omit the spin index $\bar\sigma$. By using the following identity
\bn
{\frac {\delta G(t_1,t_1^+)} {\delta\lambda_\eta(t)}}&=&-\int dt_2 dt_3 G(t_1,t_2){\frac {\delta G^{-1}(t_2,t_3)} {\delta\lambda_\eta(t)}} G(t_3,t_1^+)\nonumber\\
&=& \int dt_2 dt_3 G(t_1,t_2)[\Gamma^{(0)}_\eta(t_2,t_3,t)+\Gamma^{U}_\eta(t_2,t_3,t)]G(t_3,t_1^+)\;,
\en
 we will derive the integral equation satisfied by the interaction vertex function. If we rewrite the vertex function $\Gamma^U_\eta(t_1,t_2;t)=\tilde\Gamma^U_\eta(t_1,t)\delta(t_1,t_2)$, then the integral equation for $\tilde\Gamma^U_\eta(t_1,t)$
is given by
\bq
\tilde\Gamma^U_\eta(t_1,t)+i U\int dt_2 G(t_1,t_2)G(t_2,t_1^+)\tilde\Gamma^U_\eta(t_2,t)=-i U \Lambda^{(0)}_\eta (t_1, t)  \;,
\eq
where
\bq
\Lambda^{(0)}_\eta (t_1, t) =\int dt_2 dt_3 G(t_1, t_2)\Gamma^{(0)}_\eta(t_2,t_3,t)G(t_3,t_1^+)\;.
\eq
Introduce a function $M(t_2, t)$ which satisfies the following equation
\bq
\int dt_2 [\delta(t_1,t_2)+iU G(t_1,t_2)G(t_2,t_1^+)] M(t_2,t)=\delta(t_1, t)\;,
\eq
then we obtain the solution for the vertex function as
\bq
\tilde\Gamma^U_\eta(t_1,t)=-i U \int dt_2 M(t_1, t_2)\Lambda^{(0)}_\eta (t_2, t)\;.
\eq
It is noticed that $\Lambda^{(0)}_\eta (t_2, t)$ is equal to the  bare correlation function
of the density operator on the quantum dot $\Lambda^{(0)}_\eta (t_2, t)=\langle T_C \delta j_\eta(t)\delta n_{d\bar\sigma}(t_2) \rangle $ without the interaction vertex correction as defined in Appendix A. The function $M(t_1,t_2)$ can be obtained by solving  Eq.(B5). In the frequency space, it can be written in the matrix form of
\bq\bigg (
\begin{array}{cc}
M^{--}(\omega) & M^{-+}(\omega)\\
M^{+-}(\omega) & M^{++}(\omega)\\
\end{array}
\bigg )={\frac {1} {\Delta(\omega)}} \bigg (
\begin{array}{cc}
1+U\chi^{++,(0)}_{nn}(\omega) & U\chi^{-+,(0)}_{nn}(\omega)\\
U\chi^{+-,(0)}_{nn}(\omega) & -1+U\chi^{--,(0)}_{nn}(\omega)\\
\end{array}
\bigg )\;,
\eq
with $\Delta(\omega)=[1- \chi^{r,(0)}_{nn}(\omega)][1- \chi^{a,(0)}_{nn}(\omega)]$. Therefore,  an explicit formula Eq.(B6) for the vertex function $\tilde\Gamma^U_\eta(t_1,t)$ is obtained, which can be substituted to Eq.(32) to get the current noise interaction correction term.
\end{widetext}

\end{appendix}


\begin{thebibliography}{ }
\bibitem{Gabelli} J. Gabelli, G. F\`{e}ve, J. M. Berroir, B. Placais, A. Cavanna, B. Etienne, Y. Jin, and D. C. Glattli,
Science {\bf 313}, 499 (2006).
\bibitem{Chevallier} D. Chevallier, T. Jonckheere, E. Paladino, G. Falci, and T. Martin, Phys. Rev. B {\bf 81}, 205411 (2010).
\bibitem{Landauer} R. Landauer, Nature (London) {\bf 392}, 658 (1998)
\bibitem{Blanter} Y. M. Blanter and M. B\"uttiker, Phys. Rep. {\bf
336}, 1 (2000).
\bibitem{Levitov1993} L. S. Levitov and G. B. Lesovik, JETP Lett. {\bf
58}, 230 (1993).
\bibitem{Levitov1996} L. S. Levitov, H. W. Lee, and G. B. Lesovik, J.
Math. Phys. {\bf 37}, 4845 (1996).
\bibitem{Nazarov} Yu. V. Nazarov, Ann. Phys. (Leipzig) {\bf 8}, 507
(1999).
\bibitem{Gogolin} A. O. Gogolin and A. Komnik, Phys. Rev. B {\bf
73}, 195301 (2006); A. O. Gogolin and A. Komnik, Phys. Rev. Lett.
{\bf 97}, 016602 (2006).
\bibitem{Schmidt} T. L. Schmidt, A. Komnik, and A. O. Gogolin, Phys.
Rev. Lett. {\bf 98}, 056603 (2007).
\bibitem{Levitov2004} L. S. Levitov and M. Reznikov, Phys. Rev. B {\bf
70}, 115305 (2004).
\bibitem{Safi} I. Safi and P. Joyez, Phys. Rev. B {\bf 84}, 205129 (2011).
\bibitem{Billangeon} P. M. Billangeon, F. Pierre, H. Bouchiat, and R. Deblock, Phys. Rev. Lett. {\bf 96}, 136804 (2006).
\bibitem{Bagrets} D. A. Bagrets and Yu. V. Nazarov, Phys. Rev. B
{\bf 67}, 085316 (2003).
\bibitem{Utsumi2007} Y. Utsumi, Phys. Rev. B {\bf 75}, 035333 (2007).
\bibitem{Kambly} D. Kambly, C. Flindt, and M. B\"uttiker, Phys. Rev.
B {\bf 83}, 075432 (2011).
\bibitem{Pilgram} S. Pilgram, K. E. Nagaev, and M. B\"{u}ttiker, Phys. Rev. B {\bf 70}, 045304 (2004).
\bibitem{Lee} H. Lee, L. S. Levitov, and A. Y. Yakovets, Phy. Rev. B {\bf 51}, 4079 (1995).
\bibitem{Pedersen} M. H. Pedersen and M. B\"uttiker, Phys. Rev. B
{\bf 58}, 12993 (1998).
\bibitem{Wei} Y. D. Wei, B. G. Wang, J. Wang and H. Guo, Phys. Rev. B {\bf 60}, 16900 (1999).
\bibitem{H1992a} S. Hershfield, J. H. Davies, and J. W. Wilkins,
Phys. Rev. B {\bf 46}, 7046 (1992)
\bibitem{H1992b} S. Hershfield, Phys. Rev. B {\bf
46}, 7061 (1992).
\bibitem{Chen} L. Y. Chen and C. S. Ting, Phys. Rev. Lett. {\bf 64}, 3159 (1990); Phys. Rev. B {\bf 43}, 4534 (R) (1991).
\bibitem{Entin-Wohlman} O. Entin-Wohlman, Y. Imry, S. A. Gurvitz, and A. Aharony, Phys. Rev. B {\bf 75}, 193308 (2007).
\bibitem{Rothstein} E. A. Rothstein, O. Entin-Wohlman, and A. Aharony, Phys. Rev. B {\bf 79}, 075307 (2009).
\bibitem{Gabdank} N. Gabdank, E. A. Rothstein, O. Entin-Wohlman, and A. Aharony, Phys. Rev. B {\bf 84}, 235435 (2011).
\bibitem{Baym} G. Baym, Phys. Rev. {\bf 127}, 1391 (1962); L. P. Kadanoff and G. Baym, Phys. Rev. {\bf 124}, 287 (1961).
\bibitem{Utsumi2003} Y. Utsumi, H. Imamura, M. Hayashi, and H. Ebisawa, Phys. Rev. B {\bf 67}, 035317 (2003).
\bibitem{Oh} J. H. Oh, D. Ahn, and S. W. Hwang, Phys. Rev. B {72}, 165348 (2005).
\bibitem{Moca} C. P. Moca, P. Simon, C. H. Chung, and G. Zar\'{a}nd,  Phys. Rev. B {\bf 83}, 201303 (R) (2011).
\bibitem{Macedo} A. M. S. Mac\^{e}do, Phys. Rev. B {\bf 69}, 155309 (2004).
\bibitem{Sexty} D. Sexty, T. Gasenzer, and J. Pawlowski, Phys. Rev. B {\bf 83}, 165315 (2011).
\bibitem{Ng} T. K. Ng, Phys. Rev. Lett. {\bf 76}, 487 (1996).
\bibitem{Ding} G. H. Ding and T. K. Ng, Phys. Rev. B {\bf 56},
15521 (R) (1997).
\bibitem{Leeuwen} R. van Leeuwen and N. E. Dahlen, in {\it The Electron Liquid Mode in Condensed Matter Physics}, Vol 157 of the International School of Physics "Enrico Fermi", edited by G. F. Giuliani and G. Vignale (2004).
\bibitem{Ward} J. C. Ward, Phys Rev. B {\bf 78}, 182 (1950).
\bibitem{Langreth} D. C. Langreth, in {\it Linear and Nonlinear Electron Transport in
Solids}, Vol. 17 of Nato Advanced Study Institute, Series B:
Physics, edited by J. T. Devreese and V. E. Van Doren (Plenum,
New York, 1976).
\bibitem{Jauho} A. P. Jauho, N. S. Wingreen, and Y. Meir, Phys. Rev. B
{\bf 50}, 5528 (1994).
\bibitem{Dong} B. Dong and X. L. Lei, J. Phys. : Condens. Matter {\bf 14}, 4963 (2002).
\bibitem{Lopez} R. L\'{o}pez, R. Aguado and G. Platero, Phys. Rev. B {\bf 69}, 235305 (2004).
\bibitem{Das} M. P. Das and F. Green, J. Phys: Condens. Matter {\bf 24}, 183201 (2012).

\end{thebibliography}
\end{document}